\documentclass[11pt]{article}
\usepackage{color}
\usepackage{graphicx}
\usepackage{amsfonts}
\usepackage{theorem}
\newtheorem{Lem}{Lemma}[section]
\newtheorem{Def}[Lem]{Definition}
\newtheorem{The}[Lem]{Theorem}
\newtheorem{Prop}[Lem]{Proposition}
\newtheorem{Cor}[Lem]{Corollary}
\newtheorem{Ex}[Lem]{Example}

\newtheorem{Rem}[Lem]{Remark}

\newcommand{\qed}{\hbox{\rule{6pt}{6pt}}}

\begin{document}
\title{Schr\"odinger uncertainty relation, Wigner-Yanase-Dyson skew information and metric adjusted correlation measure}
\author{Shigeru Furuichi$^1$\footnote{E-mail:furuichi@chs.nihon-u.ac.jp} and Kenjiro Yanagi$^2$\footnote{E-mail:yanagi@yamaguchi-u.ac.jp}\\
$^1${\small Department of Computer Science and System Analysis,}\\
{\small College of Humanities and Sciences, Nihon University,}\\
{\small 3-25-40, Sakurajyousui, Setagaya-ku, Tokyo, 156-8550, Japan}\\
$^2${\small Division of Applied Mathematical Science,}\\
{\small  Graduate School of Science and Engineering, Yamaguchi University,}\\
{\small 2-16-1, Tokiwadai, Ube City, 755-0811, Japan}}
\date{}
\maketitle

{\bf Abstract.} In this paper, we give Schr\"odinger-type uncertainty relation using the Wigner-Yanase-Dyson skew information.
In addition, we give Schr\"odinger-type uncertainty relation by use of a two-parameter extended correlation measure.
We finally show the further generalization of  Schr\"odinger-type uncertainty relation by use of the metric adjusted correlation measure.
These results generalize our previous result in [Phys. Rev. A, Vol.82(2010), 034101]. 
\vspace{3mm}

\vspace{3mm}

\vspace{3mm}

%%%%%%%%%%%%%%%%%%%%%%%%%%%%%%%%%%%%%%%%%%%%%%%%%%%%%%%%%%%%%%%%%%%%%%%%%%%%%%%%%%%%%%%%%%%%%%%%%%%%%%%%%%%%%%%%%%%%%%%%%%%%%%%%%%%%%%%%%%%%%%%%%%%%%%%%%%%%%%%
%%%%%%%%%%%%%%%%%%%%%%%%%%%%%%%%%%%%%%%%%%%%%%%%%%%%%%%%%%%%%%%%%%%%%%%%%%%%%%%%%%%%%%%%%%%%%%%%%%%%%%%%%%%%%%%%%%%%%%%%%%%%%%%%%%%%%%%%%%%%%%%%%%%%%%%%%%%%%%%
%%%%%%%%%%%%%%%%%%%%%%%%%%%%%%%%%%%%%%%%%%%%%%%%%%%%%%%%%%%%%%%%%%%%%%%%%%%%%%%%%%%%%%%%%%%%%%%%%%%%%%%%%%%%%%%%%%%%%%%%%%%%%%%%%%%%%%%%%%%%%%%%%%%%%%%%%%%%%%%
%%%%%%%%%%%%%%%%%%%%%%%%%%%%%%%%%%%%%%%%%%%  Section1  %%%%%%%%%%%%%%%%%%%%%%%%%%%%%%%%%%%%%%%%%%%%%%%%%%%%%%%%%%%%%%%%%%%%%%%%%%%%%%%%%%%%%%%%%%%%%%%%%%%%%%%%
%%%%%%%%%%%%%%%%%%%%%%%%%%%%%%%%%%%%%%%%%%%%%%%%%%%%%%%%%%%%%%%%%%%%%%%%%%%%%%%%%%%%%%%%%%%%%%%%%%%%%%%%%%%%%%%%%%%%%%%%%%%%%%%%%%%%%%%%%%%%%%%%%%%%%%%%%%%%%%%
%%%%%%%%%%%%%%%%%%%%%%%%%%%%%%%%%%%%%%%%%%%%%%%%%%%%%%%%%%%%%%%%%%%%%%%%%%%%%%%%%%%%%%%%%%%%%%%%%%%%%%%%%%%%%%%%%%%%%%%%%%%%%%%%%%%%%%%%%%%%%%%%%%%%%%%%%%%%%%%
%%%%%%%%%%%%%%%%%%%%%%%%%%%%%%%%%%%%%%%%%%%%%%%%%%%%%%%%%%%%%%%%%%%%%%%%%%%%%%%%%%%%%%%%%%%%%%%%%%%%%%%%%%%%%%%%%%%%%%%%%%%%%%%%%%%%%%%%%%%%%%%%%%%%%%%%%%%%%%%
\section{Introduction}
In quantum information theory, one of the most important results is the strong subadditivity of von Neumann entropy \cite{NC}.
This important property of von Neumann entropy can be proven by the use of Lieb's theorem \cite{Lieb} which gave a complete solution for
the conjecture of the convexity of Wigner-Yanase-Dyson skew information. 
In addition, the uncertainty relation has been widely studied in quantum information theory \cite{MI,VT,TR}.
In particular,  the relations between skew information and uncertainty relation have been studied in \cite{Luo0,Furu,GII1,GII2,GHP}.
Quantum Fisher information is also called monotone metric which was introduced by Petz \cite{Petz1} 
and the Wigner-Yanase-Dyson metric is connected
to quantum Fisher information (monotone metric) as a special case. Recently, Hansen gave a further development
of the notion of monotone metric, so-called metric adjusted skew information \cite{Han}. 
The Wigner-Yanase-Dyson skew information is also connected to the metric adjusted skew information as a special case.
That is, the metric adjusted skew information gave a class including the Wigner-Yanase-Dyson skew information, while
the monotone metric gave a class including the Wigner-Yanase-Dyson metric.
In the paper  \cite{Han}, the metric adjusted correlation measure was also introduced as a generalization of the quantum covariance and correlation measure defined in
\cite{Luo0}. 
Therefore there is a significance to give the relation among the Wigner-Yanase-Dyson skew information, metric adjusted correlation measure
and uncertainty relation for the fundamental studies on quantum information theory. 

We start from the Heisenberg uncertainty relation \cite{Hei}:
\begin{equation}   \label{HUL}
V_{\rho}(A) V_{\rho}(B) \geq \frac{1}{4}\vert Tr[\rho[A,B]]\vert^2 
\end{equation}
for a quantum state (density operator) $\rho$ and two observables (self-adjoint operators) $A$ and $B$.
The further stronger result was given by Schr\"odinger in \cite{Rob,Schr}:
\begin{equation} \label{S_UL}
V_{\rho}(A) V_{\rho}(B)-\vert Re\left\{ Cov_{\rho}(A,B) \right\} \vert^2 \geq \frac{1}{4}\vert Tr[\rho[A,B]]\vert^2,  
\end{equation}
where the covariance is defined by $Cov_{\rho}(A,B) \equiv Tr[\rho \left(A-Tr[\rho A]I\right)\left(B-Tr[\rho B]I\right)].$

The Wigner-Yanase skew information represents a measure for non-commutativity between a quantum state $\rho$ and an observable $H$. 
Luo introduced the quantity $U_{\rho}(H)$ representing a quantum uncertainty excluding the classical mixture \cite{Luo1}:
\begin{equation}  \label{cla_mix}
U_{\rho}(H)  \equiv \sqrt{V_{\rho}(H)^2 -\left( V_{\rho}(H)-I_{\rho}(H)\right)^2},
\end{equation}
with the Wigner-Yanase skew information \cite{WY}:
$$
I_{\rho}(H) \equiv \frac{1}{2} Tr\left[(i[\rho^{1/2},H_0])^2\right] = Tr[\rho H_0^2] -Tr[\rho^{1/2} H_0 \rho^{1/2} H_0],\quad H_0 \equiv H-Tr[\rho H]I
$$
and then he successfully showed a new Heisenberg-type uncertainty relation on $U_{\rho}(H)$ in \cite{Luo1}:
\begin{equation}   \label{UL_U}
U_{\rho}(A)U_{\rho}(B) \geq  \frac{1}{4}\vert Tr[\rho[A,B]]\vert^2.
\end{equation}
As stated in \cite{Luo1}, the physical meaning of the quantity $U_{\rho}(H)$ can be interpreted as follows.
For a mixed state $\rho$, the variance $V_{\rho}(H)$ has both classical mixture and quantum uncertainty.
Also, the Wigner-Yanase skew information $I_{\rho}(H)$ represents a kind of quantum uncertainty \cite{Luo2,Luo3}. Thus, the difference
$V_{\rho}(H) - I_{\rho}(H)$ has a classical mixture so that we can regard that the quantity $U_{\rho}(H)$  
has a quantum uncertainty excluding a classical mixture. Therefore it is meaningful and suitable to study an uncertainty
relation for a mixed state by the use of  the quantity $U_{\rho}(H)$.

Recently, a one-parameter extension of the inequality (\ref{UL_U}) was given in \cite{Yanagi1}:
\begin{equation}   \label{gen_UL_U}
U_{\rho,\alpha}(A)U_{\rho,\alpha}(B) \geq \alpha (1-\alpha) |Tr[\rho[A,B]]|^2,
\end{equation}
where
$$
U_{\rho,\alpha}(H)  \equiv \sqrt{V_{\rho}(H)^2 -\left( V_{\rho}(H)-I_{\rho,\alpha}(H)\right)^2},
$$
with the Wigner-Yanase-Dyson skew information $I_{\rho,\alpha}(H)$ is defined by
$$
I_{\rho,\alpha}(H) \equiv \frac{1}{2} Tr\left[(i[\rho^{\alpha},H_0])(i[\rho^{1-\alpha},H_0])\right] = Tr[\rho H_0^2] -Tr[\rho^{\alpha} H_0 \rho^{1-\alpha} H_0],
$$
It is notable that the convexity of $I_{\rho,\alpha}(H)$ with respect to $\rho$ was successfully proven by Lieb in \cite{Lieb}.
The further generalization of the  Heisenberg-type uncertainty relation on $U_{\rho}(H)$ has been given in \cite{Yanagi2} using
the generalized Wigner-Yanase-Dyson skew information introduced in \cite{CL}.
See also \cite{ACH,FYK,GHP,GII1} for the recent studies on skew informations and uncertainty relations.

Motivated by the fact that the Schr\"odinger uncertainty relation is a stronger result than the Heisenberg uncertainty relation,
  a new Schr\"odinger-type uncertainty relation for mixed states using Wigner-Yanase skew information was shown in \cite{Furu}.
That is, for a quantum state  $\rho$ and two observables $A$ and $B$, we have
\begin{equation} \label{sul_wy}
U_\rho(A)U_\rho(B)-|Re\left\{ Corr_{\rho}(A,B)\right\}|^2 \geq  \frac{1}{4} |Tr[\rho[A,B]]|^2,
\end{equation}
where the correlation measure \cite{Luo0} is defined by
$$
Corr_{\rho}(X,Y) \equiv Tr[\rho X^*Y] -Tr[\rho^{1/2} X^*\rho^{1/2}Y]
$$
for any operators $X$ and $Y$. 
This result refined the Heisenberg-type uncertainty relation (\ref{UL_U}) shown in \cite{Luo1}  for mixed states (general states).
We easily find  that the inequality (\ref{sul_wy}) is equivalent to the following inequality:
\begin{equation} \label{sul_wy02}
U_\rho(A)U_\rho(B) \geq | Corr_{\rho}(A,B) |^2. 
\end{equation}
The main purpose of this paper is to give some extensions of the inequality (\ref{sul_wy02}) 
by using the Wigner-Yanase-Dyson skew information $I_{\rho,\alpha}(H)$ and the metric adjusted correlation measure introduced in \cite{Han}. 

%%%%%%%%%%%%%%%%%%%%%%%%%%%%%%%%%%%%%%%%%%%%%%%%%%%%%%%%%%%%%%%%%%%%%%%%%%%%%%%%%%%%%%%%%%%%%%%%%%%%%%%%%%%%%%%%%%%%%%%%%%%%%%%%%%%%%%%%%%%%
%%%%%%%%%%%%%%%%%%%%%%%%%%%%%%%%%%%%%%%%%%%%%%%%%%%%%%%%%%%%%%%%%%%%%%%%%%%%%%%%%%%%%%%%%%%%%%%%%%%%%%%%%%%%%%%%%%%%%%%%%%%%%%%%%%%%%%%%%%
%%%%%%%%%%%%%%%%%%%%%%%%%%%%%%%%%%%%%%%%%%%%%%%%%%%%%%%%%%%%%%%%%%%%%%%%%%%%%%%%%%%%%%%%%%%%%%%%%%%%%%%%%%%%%%%%%%%%%%%%%%%%%%%%%%%%%%%%%%
%%%%%%%%%%%%%%%%%%%%%%%%%%%%%%%%%%%%%%%%%%%%%%%%%%%%%%%%%%%%%%%%%%%%%%%%%%%%%%%%%%%%%%%%%%%%%%%%%%%%%%%%%%%%%%%%%%%%%%%%%%%%%%%%%%%%%%%%
%%%%%%%%%%%%%%%%%%%%%%%%%%%%%%%%%%%%%%%%%%%%%%%%%%%%%%%%%%%%%%%%%%%%%%%%%%%%%%%%%%%%%%%%%%%%%%%%%%%%%%%%%%%%%%%%%%%%%%%%%%%%%%%%%%%%%%%%
%%%%%%%%%%%%%%%%%%%%%%%%%%%%%%%%%%%%%%%%%%%%%%%%%%%%%%%%%%%%%%%%%%%%%%%%%%%%%%%%%%%%%%%%%%%%%%%%%%%%%%%%%%%%%%%%%%%%%%%%%%%%%%%%%%%%%%%%%%
\section{Schr\"odinger uncertainty relation with Wigner-Yanase-Dyson skew information}

In this section, we give a generalization of the  Schr\"odinger type uncertainty relation (\ref{sul_wy02})  by the use of the quantity
$U_{\rho,\alpha}(H)$ defined by the Wigner-Yanase-Dyson skew information $I_{\rho,\alpha}(H)$.

\begin{The}   \label{the_1_2}
For $\alpha \in [1/2,1]$, a quantum state $\rho$ and two observables $A$ and $B$, we have
\begin{equation} \label{ineq_GSUL}
U_{\rho,\alpha}(A)U_{\rho,\alpha}(B) \geq 4 \alpha (1-\alpha) \vert Corr_{\rho,\alpha}(A,B) \vert^2.
\end{equation} 
where the generalized correlation measure \cite{Kos,YFK} is defined by
$$
Corr_{\rho,\alpha}(X,Y) \equiv Tr[\rho X^*Y] -Tr[\rho^{\alpha} X^*\rho^{1-\alpha}Y]
$$
for any operators $X$ and $Y$. 
\end{The}

To prove Theorem \ref{the_1_2}, we need the following lemmas.
\begin{Lem}  {\bf (\cite{Yanagi1})}   \label{lem-1-2-01}
For a spectral decomposition of $\rho = \sum_{j=1}^{\infty} \lambda_j \vert \phi_j \rangle \langle \phi_j \vert$, 
putting $h_{ij} \equiv \langle \phi_i \vert H_0 \vert \phi_j \rangle$, we have the following relations.
\begin{itemize}
\item[(i)] For the Wigner-Yanase-Dyson skew information, we have
$$
I_{\rho,\alpha} (H) = \sum_{i<j} \left(\lambda_i^{\alpha} - \lambda_j^{\alpha}\right)  \left(\lambda_i^{1-\alpha}- \lambda_j^{1-\alpha}\right) \vert h_{ij}\vert^2.
$$
\item[(ii)]For the quantity associated to the Wigner-Yanase-Dyson skew information:
$$
J_{\rho,\alpha}(H) \equiv \frac{1}{2} Tr\left[\left(\left\{\rho^{\alpha},H_0\right\}\right)(\left\{\rho^{1-\alpha},H_0\right\})\right] =
 Tr[\rho H_0^2] +Tr[\rho^{\alpha} H_0 \rho^{1-\alpha} H_0],
$$
where $\left\{X,Y\right\} \equiv XY+YX$ is an anti-commutator,
we have
$$
J_{\rho,\alpha} (H) \geq \sum_{i<j} \left(\lambda_i^{\alpha} + \lambda_j^{\alpha}\right)  \left(\lambda_i^{1-\alpha}+ \lambda_j^{1-\alpha}\right) \vert h_{ij}\vert^2.
$$
\end{itemize}
\end{Lem}

\begin{Lem}{\bf (\cite{BD,Yanagi1})}   \label{lem-1-2-02}
For any $t>0$ and $\alpha \in [0,1]$, we have
$$
(1-2\alpha)^2(t-1)^2 \geq (t^{\alpha}-t^{1-\alpha})^2.
$$
\end{Lem}

{\it Proof of Theorem \ref{the_1_2}}:
We take a spectral decomposition $\rho = \sum_{j=1}^{\infty} \lambda_j \vert \phi_j \rangle \langle \phi_j \vert$.
If we put $a_{ij}=  \langle \phi_i\vert A_0 \vert \phi_j \rangle$ and $b_{ji}= \langle \phi_j\vert B_0 \vert \phi_i \rangle$,
where $A_0=A-Tr[\rho A] I$ and $B_0=B-Tr[\rho B] I$, then we have
\begin{eqnarray}
Corr_{\rho, \alpha}(A,B) &=& Tr[\rho A B] -Tr[\rho^{\alpha} A \rho^{1-\alpha} B] \nonumber \\
&=& Tr[\rho A_0 B_0] -Tr[\rho^{\alpha} A_0 \rho^{1-\alpha} B_0] \nonumber \\
&=& \sum_{i,j=1}^{\infty}(\lambda_i-\lambda_i^{\alpha}\lambda_j^{1-\alpha}) a_{ij} b_{ji}\nonumber \\
&=& \sum_{i\neq j} (\lambda_i-\lambda_i^{\alpha}\lambda_j^{1-\alpha}) a_{ij} b_{ji}\nonumber \\
& =& \sum_{i < j}  \left\{ (\lambda_i-\lambda_i^{\alpha}\lambda_j^{1-\alpha}) a_{ij} b_{ji} + (\lambda_j-\lambda_j^{\alpha}\lambda_i^{1-\alpha}) a_{ji} b_{ij} \right\}.\label{corr_eq_1}
\end{eqnarray}
Thus we have
$$
\vert Corr_{\rho,\alpha}(A,B)\vert  \leq \sum_{i < j} \left\{  \vert \lambda_i -\lambda_i^{\alpha} \lambda_j^{1-\alpha} \vert \vert a_{ij} \vert \vert b_{ji} \vert
+  \vert \lambda_j -\lambda_j^{ \alpha} \lambda_i^{1-\alpha} \vert \vert a_{ji} \vert \vert b_{ij} \vert \right\}.
$$
Since $\vert a_{ij} \vert = \vert a_{ji} \vert$ and $\vert b_{ij} \vert = \vert b_{ji} \vert$, 
taking a square of both sides and then using Schwarz inequality and  Lemma \ref{lem-1-2-01}, we have
\begin{eqnarray*}
&& 4 \alpha(1-\alpha) \vert Corr_{\rho,\alpha}(A,B)\vert ^2 \\
&&\leq  4 \alpha(1-\alpha) \left\{ \sum_{i < j}  \left\{  \vert \lambda_i -\lambda_i^{\alpha} \lambda_j^{1-\alpha} \vert 
+  \vert \lambda_j -\lambda_j^{\alpha} \lambda_i^{1-\alpha} \vert \right\} \vert a_{ij} \vert \vert b_{ji} \vert \right\}^2 \\
&&= \left\{ \sum_{i<j} 2\sqrt{\alpha(1-\alpha)}\left( \lambda_i^{\alpha}+\lambda_{j}^{\alpha} \right) \vert \lambda_i^{1-\alpha} -\lambda_j^{1-\alpha} \vert \vert a_{ij} \vert \vert b_{ji} \vert \right\}^2 \\
&&\leq  \left\{ \sum_{i<j} 2\sqrt{\alpha(1-\alpha)} \vert \lambda_i - \lambda_j \vert  \vert a_{ij} \vert \vert b_{ji} \vert \right\}^2 \\
&&\leq  \left\{ \sum_{i<j}  
\left\{ \left( \lambda_i^{\alpha} - \lambda_j^{\alpha} \right) \left( \lambda_i^{1-\alpha} - \lambda_j^{1-\alpha} \right) 
\left( \lambda_i^{\alpha} + \lambda_j^{\alpha} \right) \left( \lambda_i^{1-\alpha} + \lambda_j^{1-\alpha} \right) \right\}^{1/2}  \vert a_{ij} \vert \vert b_{ji} \vert \right\}^2 \\
&& \leq  \left\{\sum_{i<j} \left( \lambda_i^{\alpha} - \lambda_j^{\alpha} \right) \left( \lambda_i^{1-\alpha} - \lambda_j^{1-\alpha} \right) \vert a_{ij} \vert^2\right\} 
\left\{\sum_{i<j} \left( \lambda_i^{\alpha} + \lambda_j^{\alpha} \right) \left( \lambda_i^{1-\alpha} + \lambda_j^{1-\alpha} \right) \vert b_{ij} \vert^2\right\} \\
&& \leq  I_{\rho,\alpha}(A) J_{\rho,\alpha}(B)
\end{eqnarray*}
In the above process, the inequality 
$(x^{\alpha}+y^{\alpha}) \vert x^{1- \alpha} -y^{1-\alpha} \vert \leq \vert x-y \vert $ for $x,y \geq 0 $ and $\alpha \in [\frac{1}{2},1]$
and the inequality $4 \alpha (1-\alpha)   (x-y)^2 \leq  \left( x^{\alpha} - y^{\alpha} \right) \left( x^{1-\alpha} - y^{1-\alpha} \right)
 \left( x^{\alpha} + y^{\alpha} \right) \left( x^{1-\alpha} + y^{1-\alpha} \right)$ for $x,y \geq 0 $ and $\alpha \in [0,1]$,
which can be proven by Lemma \ref{lem-1-2-02}, were used.
By the similar way, we also have
$$
4 \alpha(1-\alpha) \vert Corr_{\rho,\alpha}(A,B)\vert ^2 \leq I_{\rho,\alpha}(B) J_{\rho,\alpha}(A).
$$
Thus for $\alpha \geq \frac{1}{2}$ we have
\begin{equation}
4 \alpha(1-\alpha) \vert Corr_{\rho,\alpha}(A,B)\vert ^2 \leq U_{\rho,\alpha}(A) U_{\rho,\alpha}(B).
\end{equation}

\hfill \qed

Note that Theorem \ref{the_1_2} recovers the inequality (\ref{sul_wy02}), if we take $\alpha = \frac{1}{2}$. 
%We also note that the following inequality does not hold in general,
%$$2\sqrt{\alpha(1-\alpha)}\left( x^{\alpha}+y^{\alpha} \right) \vert x^{1-\alpha} -y^{1-\alpha} \vert  \leq \left\{ \left( x^{\alpha} - y^{\alpha} \right) \left( x^{1-\alpha} - y^{1-\alpha} \right)  \left( x^{\alpha} + y^{\alpha} \right) \left( x^{1-\alpha} + y^{1-\alpha} \right) \right\}^{1/2}$$
%for $x,y, \alpha \in [0,1]$. 

\begin{Rem}
We take $\alpha = 0.1$ and 
\[
\rho  = \frac{1}{3}\left( {\begin{array}{*{20}c}
   1 & 0  \\
   0 & 2  \\
\end{array}} \right),A = \left( {\begin{array}{*{20}c}
   2 & {2 - i}  \\
   {2 + i} & 1  \\
\end{array}} \right),B = \left( {\begin{array}{*{20}c}
   2 & i  \\
   { - i} & 1  \\
\end{array}} \right),
\]
then we have 
$$
U_{\rho,\alpha}(A)U_{\rho,\alpha}(B) - 4 \alpha (1-\alpha) \vert Corr_{\rho,\alpha}(A,B) \vert^2 \simeq -0.28332.
$$
Therefore the inequality (\ref{ineq_GSUL}) does not hold for $\alpha \in [0,1/2)$ in general.

%\textcolor{red}{We note that from Lemma \ref{lem-1-2-01}, we find that $I_{\rho,\alpha}(H) = I_{\rho,1-\alpha}(H)$ and 
%$$J_{\rho,1-\alpha}(H) \geq \sum_{i<j} \left(\lambda_i^{\alpha} + \lambda_j^{\alpha}\right)  \left(\lambda_i^{1-\alpha}+ \lambda_j^{1-\alpha}\right) \vert h_{ij}\vert^2.$$
%However the example given in Remark \ref{rem_corr_tr} shows $\vert Corr_{\rho,\alpha}(A,B)\vert \simeq 1.38821$ and $\vert Corr_{\rho,1-\alpha}(A,B)\vert \simeq 2.08166$ so that $\vert Corr_{\rho,\alpha}(A,B) \vert \neq \vert Corr_{\rho,1-\alpha}(A,B)\vert $ in general.}
\end{Rem}

\begin{Cor}
Under the same assumptions with Theorem \ref{the_1_2}, 
we have the following inequality:
\begin{eqnarray}
&&U_{\rho,\alpha}(A) U_{\rho,\alpha}(B) -4 \alpha (1-\alpha) \left( \vert Re\left\{Corr_{\rho,\alpha}(A,B)\right\}\vert^2
- \vert Im \left\{  Tr[\rho^{\alpha}A\rho^{1-\alpha}B] \right\} \vert^2 \right) \nonumber \\
&& \hspace*{3cm} \geq \alpha (1-\alpha) \vert Tr[\rho[A,B]]\vert^2.   \label{cor_alpha}
\end{eqnarray}
\end{Cor}
{\it Proof}:
From 
$$
Im\left\{ Corr_{\rho,\alpha}(A,B)\right\} = \frac{1}{2i} Tr\left[\rho[A,B]\right] - Im\left\{Tr[\rho^{\alpha} A\rho^{1-\alpha} B ]\right\},
$$
we have
$$
\frac{1}{4} \vert Tr\left[ \rho[A,B]\right]\vert^2 \leq \vert Im\left\{ Corr_{\rho,\alpha}(A,B) \right\} \vert^2 + 
\vert Im\left\{Tr[\rho^{\alpha}A\rho^{1-\alpha}B]\right\}\vert^2.
$$
Thus we have
\begin{eqnarray*}
\vert Corr_{\rho,\alpha}(A,B)\vert^2 &=& \vert Re\left\{ Corr_{\rho,\alpha}(A,B) \right\} \vert^2 + \vert Im\left\{ Corr_{\rho,\alpha}(A,B) \right\} \vert^2 \\
&\geq &  \vert Re\left\{ Corr_{\rho,\alpha}(A,B) \right\} \vert^2 +\frac{1}{4} \vert Tr\left[ \rho[A,B]\right]\vert^2 -\vert Im\left\{Tr[\rho^{\alpha}A\rho^{1-\alpha}B]\right\}\vert^2,
\end{eqnarray*}
which proves the corollary.

\hfill \qed

\begin{Rem}  \label{rem_corr_tr}
The following inequality does not hold in general for $\alpha\in [\frac{1}{2},1]$:
\begin{equation}
\vert Re\left\{Corr_{\rho,\alpha}(A,B)\right\}\vert^2 \geq \vert Im \left\{  Tr[\rho^{\alpha}A\rho^{1-\alpha}B] \right\} \vert^2. 
\end{equation}
Because we have a counter-example as follows.
We take $\alpha = \frac{2}{3}$ and 
\[
\rho  = \frac{1}{7}\left( {\begin{array}{*{20}c}
   2 & 3  \\
   3 & 5  \\
\end{array}} \right),A = \left( {\begin{array}{*{20}c}
   2 & {2 - i}  \\
   {2 + i} & 1  \\
\end{array}} \right),B = \left( {\begin{array}{*{20}c}
   2 & i  \\
   { - i} & 1  \\
\end{array}} \right),
\]
then we have 
$$
\vert Re\left\{Corr_{\rho,\alpha}(A,B)\right\}\vert^2 -\vert Im \left\{  Tr[\rho^{\alpha}A\rho^{1-\alpha}B] \right\} \vert^2 \simeq -0.0548142. 
$$
This  shows Theorem \ref{the_1_2} does not refine the inequality (\ref{gen_UL_U}) in general.
\end{Rem}

%%%%%%%%%%%%%%%%%%%%%%%%%%%%%%%%%%%%%%%%%%%%%%%%%%%%%%%%%%%%%%%%%%%%%%%%%%%%%%%%%%%%%%%%%%%%%%%%%%%%%%%%%%%%%%%%%%%%%%%%%%%%%%%%%%%%%%%%%%%%
%%%%%%%%%%%%%%%%%%%%%%%%%%%%%%%%%%%%%%%%%%%%%%%%%%%%%%%%%%%%%%%%%%%%%%%%%%%%%%%%%%%%%%%%%%%%%%%%%%%%%%%%%%%%%%%%%%%%%%%%%%%%%%%%%%%%%%%%%%
%%%%%%%%%%%%%%%%%%%%%%%%%%%%%%%%%%%%%%%%%%%%%%%%%%%%%%%%%%%%%%%%%%%%%%%%%%%%%%%%%%%%%%%%%%%%%%%%%%%%%%%%%%%%%%%%%%%%%%%%%%%%%%%%%%%%%%%%%%
%%%%%%%%%%%%%%%%%%%%%%%%%%%%%%%%%%%%%%%%%%%%%%%%%%%%%%%%%%%%%%%%%%%%%%%%%%%%%%%%%%%%%%%%%%%%%%%%%%%%%%%%%%%%%%%%%%%%%%%%%%%%%%%%%%%%%%%%
%%%%%%%%%%%%%%%%%%%%%%%%%%%%%%%%%%%%%%%%%%%%%%%%%%%%%%%%%%%%%%%%%%%%%%%%%%%%%%%%%%%%%%%%%%%%%%%%%%%%%%%%%%%%%%%%%%%%%%%%%%%%%%%%%%%%%%%%
%%%%%%%%%%%%%%%%%%%%%%%%%%%%%%%%%%%%%%%%%%%%%%%%%%%%%%%%%%%%%%%%%%%%%%%%%%%%%%%%%%%%%%%%%%%%%%%%%%%%%%%%%%%%%%%%%%%%%%%%%%%%%%%%%%%%%%%%%%

\section{Two-parameter extensions}
In this section, we introduce the parametric extended correlation measure $Corr_{\rho,\alpha,\gamma}(X,Y)$
 by the convex combination between $Corr_{\rho,\alpha}(X,Y) $ and $Corr_{\rho,1-\alpha}(X,Y)$.
Then we establish the parametric extended Schr\"odinger-type uncertainty relation applying the parametric extended correlation measure $Corr_{\rho,\alpha,\gamma}(X,Y)$.
 In addition,  introducing the symmetric extended correlation measure $Corr^{(sym)}_{\rho,\alpha,\gamma}(X,Y)$
 by the convex combination between $Corr_{\rho,\alpha}(X,Y) $ and $Corr_{\rho,\alpha}(Y,X)$, we show its Schr\"odinger-type uncertainty relation.

\begin{Def} \label{gen_CM}
We define the parametric extended correlation measure $Corr_{\rho,\alpha,\gamma}(X,Y)$ for two parameters $\alpha,\gamma \in [0,1]$ by
\begin{equation} 
Corr_{\rho,\alpha,\gamma}(X,Y) \equiv \gamma Corr_{\rho,\alpha}(X,Y) + (1-\gamma) Corr_{\rho,1-\alpha}(X,Y)
\end{equation}
for any operators $X$ and $Y$.
\end{Def}
Note that we have $Corr_{\rho,\alpha,\gamma}(H,H)=I_{\rho,\alpha}(H)$ for any observable $H$.
Then we can prove the following inequality.

\begin{The} \label{the_1_3}
If $0 \leq \alpha, \gamma \leq \frac{1}{2} $ or $\frac{1}{2} \leq \alpha, \gamma \leq 1$, then we have
$$
U_{\rho,\alpha}(A)  U_{\rho,\alpha}(B) \geq 4\alpha (1-\alpha) \vert Corr_{\rho,\alpha,\gamma}(A,B)\vert^2  
$$
 for two observables $A$, $B$ and a quantum state $\rho$.
\end{The}

{\it Proof}:
By the similar way of the proof of Theorem \ref{the_1_2}, we have Eq.(\ref{corr_eq_1}) and we also have
\begin{eqnarray}
Corr_{\rho, 1-\alpha}(A,B) &=& Tr[\rho A B] -Tr[\rho^{1-\alpha} A \rho^{\alpha} B] \nonumber \\
& =& \sum_{i < j}  \left\{ (\lambda_i-\lambda_i^{1-\alpha}\lambda_j^{\alpha}) a_{ij} b_{ji} + (\lambda_j-\lambda_j^{1-\alpha}\lambda_i^{\alpha}) a_{ji} b_{ij} \right\}.\label{corr_eq_2}
\end{eqnarray}
Thus we have
\begin{eqnarray*}
Corr_{\rho, \alpha,\gamma}(A,B) &=& \gamma Corr_{\rho, \alpha}(A,B) + (1-\gamma) Corr_{\rho, \alpha}(A,B) \\
&=&\sum_{i<j} \left\{ \gamma \lambda_i^{\alpha}(\lambda_i^{1-\alpha}-\lambda_j^{1-\alpha}) +(1-\gamma) \lambda_i^{1-\alpha}(\lambda_i^{\alpha}-\lambda_j^{\alpha})\right\} a_{ij}b_{ji} \\
&&\quad + \sum_{i<j} \left\{  \gamma \lambda_j^{\alpha}(\lambda_j^{1-\alpha}-\lambda_i^{1-\alpha}) 
+(1-\gamma) \lambda_j^{1-\alpha}(\lambda_j^{\alpha}-\lambda_i^{\alpha}) \right\} a_{ji}b_{ij}.
\end{eqnarray*}
Since we have $\vert a_{ij} \vert = \vert a_{ji} \vert$ and $\vert b_{ij} \vert = \vert b_{ji} \vert$, we then have
\begin{eqnarray*}
\vert Corr_{\rho, \alpha,\gamma}(A,B) \vert &\leq& \sum_{i<j} \left\{ \gamma (\lambda_i^{\alpha} + 
\lambda_j^{\alpha})\vert \lambda_i^{1-\alpha}-\lambda_j^{1-\alpha}\vert +(1-\gamma) (\lambda_i^{1-\alpha}+\lambda_j^{1-\alpha})\vert \lambda_i^{\alpha}-\lambda_j^{\alpha}\vert \right\}   \vert a_{ij}\vert \vert b_{ji}\vert   \\
&\leq & \sum_{i<j} \vert \lambda_i-\lambda_j\vert \vert a_{ij}\vert \vert b_{ji}\vert,   
\end{eqnarray*}
thanks to the inequality
\begin{equation} \label{ineq_eq}
\gamma (x^{\alpha}+y^{\alpha})\vert x^{1-\alpha}-y^{1-\alpha} \vert + (1-\gamma)  (x^{1-\alpha}+y^{1-\alpha})\vert x^{\alpha}-y^{\alpha} \vert \leq \vert x-y \vert
\end{equation}
for $0 \leq \alpha, \gamma \leq \frac{1}{2} $ or $\frac{1}{2} \leq \alpha, \gamma \leq 1$, and $x,y \geq 0$.
The rest of the proof goes similar way to that of Theorem \ref{the_1_2}.

\hfill \qed

\begin{Cor}  \label{cor_two_para}
For any $\alpha \in [0,1]$, two observables $A$, $B$ and a quantum state $\rho$, we have
$$
U_{\rho,\alpha}(A)  U_{\rho,\alpha}(B) \geq 4\alpha (1-\alpha) \vert Corr_{\rho,\alpha,\frac{1}{2}}(A,B)\vert^2.  
$$
%where we call $Corr_{\rho,\alpha,\frac{1}{2}}(A,B)$ a symmetrized correlation measure.
\end{Cor}

{\it Proof}:
If $\gamma = \frac{1}{2}$, then the equality of the inequality (\ref{ineq_eq}) holds for any $\alpha \in [0,1]$ and $x,y \geq 0$.
Therefore we have the present corollary from Theorem \ref{the_1_3}.
\hfill \qed

We may define the following correlation measure instead of Definition \ref{gen_CM}.
\begin{Def}
We define a symmetric extended correlation measure $Corr^{(sym)}_{\rho,\alpha,\gamma}(X,Y)$ for two parameters $\alpha,\gamma \in [0,1]$ by
\begin{equation} 
Corr^{(sym)}_{\rho,\alpha,\gamma}(X,Y) \equiv \gamma Corr_{\rho,\alpha}(X,Y) + (1-\gamma) Corr_{\rho,\alpha}(Y,X)
\end{equation}
for any operators $X$ and $Y$.
\end{Def}

Note that we have $Corr^{(sym)}_{\rho,\alpha,\gamma}(A,B) = Corr^{(sym)}_{\rho,\alpha,\gamma}(B,A)$ for self-adjoint operators $A$ and $B$.
Then we have the following therem by the similar proof of the above using the inequality
$$(x^{\alpha}+y^{\alpha})\vert x^{1-\alpha} -y^{1-\alpha}\vert \leq \vert x -y \vert $$
for $x,y \geq 0$ and $\alpha \geq \frac{1}{2}$. 

\begin{The} \label{the_1_3_2}
For $\alpha \in [\frac{1}{2},1]$ and $\gamma \in [0,1]$, we have
$$
U_{\rho,\alpha}(A)  U_{\rho,\alpha}(B) \geq 4\alpha (1-\alpha) \vert Corr^{(sym)}_{\rho,\alpha,\gamma}(A,B)\vert^2  
$$
 for two observables $A$, $B$ and a quantum state $\rho$.
\end{The}

%%%%%%%%%%%%%%%%%%%%%%%%%%%%%%%%%%%%%%%%%%%%%%%%%%%%%%%%%%%%%%%%%%%%%%%%%%%%%%%%%%%%%%%%%%%%%%%%%%%%%%%%%%%%%%%%%%%%%%%%%%%%%%%%%%%%%%%%%%%%
%%%%%%%%%%%%%%%%%%%%%%%%%%%%%%%%%%%%%%%%%%%%%%%%%%%%%%%%%%%%%%%%%%%%%%%%%%%%%%%%%%%%%%%%%%%%%%%%%%%%%%%%%%%%%%%%%%%%%%%%%%%%%%%%%%%%%%%%%%
%%%%%%%%%%%%%%%%%%%%%%%%%%%%%%%%%%%%%%%%%%%%%%%%%%%%%%%%%%%%%%%%%%%%%%%%%%%%%%%%%%%%%%%%%%%%%%%%%%%%%%%%%%%%%%%%%%%%%%%%%%%%%%%%%%%%%%%%%%
%%%%%%%%%%%%%%%%%%%%%%%%%%%%%%%%%%%%%%%%%%%%%%%%%%%%%%%%%%%%%%%%%%%%%%%%%%%%%%%%%%%%%%%%%%%%%%%%%%%%%%%%%%%%%%%%%%%%%%%%%%%%%%%%%%%%%%%%
%%%%%%%%%%%%%%%%%%%%%%%%%%%%%%%%%%%%%%%%%%%%%%%%%%%%%%%%%%%%%%%%%%%%%%%%%%%%%%%%%%%%%%%%%%%%%%%%%%%%%%%%%%%%%%%%%%%%%%%%%%%%%%%%%%%%%%%%
%%%%%%%%%%%%%%%%%%%%%%%%%%%%%%%%%%%%%%%%%%%%%%%%%%%%%%%%%%%%%%%%%%%%%%%%%%%%%%%%%%%%%%%%%%%%%%%%%%%%%%%%%%%%%%%%%%%%%%%%%%%%%%%%%%%%%%%%%%

\section{A further generalization by metric adjusted correlation measure}
Inspired by the recent results in \cite{GI} and  the concept of metric adjusted skew information introduced by Hansen in \cite{Han},
we here give a further generalization for Schr\"odinger-type uncertainty relation applying metric adjusted correlation measure introduced in \cite{Han}.
We firstly give some notations according to those in \cite{GI}.
Let $M_n(\mathbb{C})$ and $M_{n,sa}(\mathbb{C})$ be the set of all $n \times n$ 
complex matrices and all $n \times n$ self-adjoint matrices, equipped with the Hilbert-Schmidt 
scalar product $\langle A,B \rangle = Tr[A^*B]$, respectively.  Let $M_{n,+}(\mathbb{C})$ be the set of all positive definite matrices of $M_{n,sa}(\mathbb{C})$ 
and $M_{n,+,1}(\mathbb{C})$ be the set of all density matrices, that is 
$$M_{n,+,1}(\mathbb{C})  \equiv \{ \rho \in M_{n,sa}(\mathbb{C}) | Tr \rho = 1, \rho > 0 \} \subset M_{n,+}(\mathbb{C}).$$ 
Here $X\in M_{n,+}(\mathbb{C})$ means we have $\langle \phi \vert X \vert \phi \rangle \geq 0$ for any vector
$\vert \phi \rangle \in \mathbb{C}^n$. In the study of quantum physics, we usually use a positive semidefinite matrix with a unit trace as a density operator $\rho$.
In this section, we assume the invertibility of $\rho$.
%If it is not otherwise specified, from now on we shall treat the case of faithful states, that is $\rho > 0$. 

A function $f:(0,+\infty) \rightarrow \mathbb{R}$ is said operator monotone if the inequalities $0 \leq f(A) \leq f(B)$ hold for any 
$A, B \in M_{n,sa}(\mathbb{C})$ such that $0 \leq A \leq B$. 
An operator monotone function $f:  (0,+\infty) \rightarrow (0,+\infty) $ is said symmetric if $f(x) = xf(x^{-1})$ and normalized if $f(1) = 1$. 
We represents the set of all symmetric normalized operator monotone functions by ${\cal F}_{op}$. 
We have the following examples as elements of ${\cal F}_{op}$: 
\begin{Ex} {\bf (\cite{Han,GI,GHI,PG})}
$$
f_{RLD}(x) = \frac{2x}{x+1}, \; \; \; f_{SLD}(x) = \frac{x+1}{2}, \; \; \; f_{BKM}(x) = \frac{x-1}{\log x},
$$
$$
 f_{WY}(x) = \left( \frac{\sqrt{x}+1}{2} \right)^2, \; \;  f_{WYD}(x) = \alpha (1-\alpha) \frac{(x-1)^2}{(x^{\alpha}-1)(x^{1-\alpha}-1)}, \; \alpha \in (0,1).
$$
\label{ex:example1}
\end{Ex}
The functions $f_{BKM}(x)$ and $f_{WYD}(x)$ are normalized in the sense that $\lim_{x\to 1} f_{BKM}(x) =1$ and $\lim_{x\to 1} f_{WYD}(x) =1$. 
Note that a simple proof of the operator monotonicity of $f_{WYD}(x)$ was given in \cite{GHI}.
See also \cite{Uch} for the proof of the operator monotonicity of $f_{WYD}(x)$ by use of majorization.
%In the rest of this section, we make a restriction on $\alpha$  in $(0,1/2]$ or $[1/2,1)$ for the function  $f_{WYD}(x) $ above.

\begin{Rem} {\bf (\cite{GI,KA,Petz3,PG})}
For any $f \in {\cal F}_{op}$, we have the following inequalities:
$$
\frac{2x}{x+1} \leq f(x) \leq \frac{x+1}{2}, \; \; x > 0.
$$
That is, all $f \in {\cal F}_{op}$ lies in between the harmonic mean and the arithmetic mean.
\label{re:remark1}
\end{Rem}

For $f \in {\cal F}_{op}$ we define $f(0) = \lim_{x \to 0}f(x)$. We also denote the sets of regular and non-regular 
functions by
$$
{\cal F}_{op}^r = \{ f \in {\cal F}_{op} | f(0) \neq 0 \} \,\,\, and \,\,\, {\cal F}_{op}^n =\{ f \in {\cal F}_{op} | f(0) = 0 \}.
$$
%and then we have ${\cal F}_{op} = {\cal F}_{op}^r \cup {\cal F}_{op}^n$. 

\begin{Def} {\bf (\cite{GII1,GI})}
For $f \in {\cal F}_{op}^r$, we define the function $\tilde{f}$ by 
$$
\tilde{f}(x) = \frac{1}{2} \left\{(x+1)-(x-1)^2 \frac{f(0)}{f(x)} \right\}, \; \; (x > 0).
$$
\label{def:definition2}
\end{Def}

Then we have the following theorem.
\begin{The}{\bf (\cite{GII1,GHI,PS})} \label{th:theorem1}
The correspondence $f \rightarrow \tilde{f}$ is a bijection between ${\cal F}_{op}^r$ and ${\cal F}_{op}^n$. 
\end{The}

We can use matrix mean theory introduced by Kubo-Ando in \cite{KA}.
Then a mean $m_f$ corresponds to each operator monotone function $f \in {\cal F}_{op}$ 
by the following formula 
$$
m_f(A,B) = A^{1/2}f(A^{-1/2}BA^{-1/2})A^{1/2}, 
$$
for $A, B \in M_{n,+}(\mathbb{C})$. By the notion of matrix mean, we may define the set of the monotone metrics \cite{Petz1}
 by the following formula 
$$
\langle A,B \rangle_{\rho,f} = Tr[A m_f(L_{\rho},R_{\rho})^{-1}(B)], 
$$
where $L_{\rho}(A) = \rho A$ and $R_{\rho}(A) = A \rho$. 

\begin{Def} {\bf (\cite{Han,GII1})}  \label{def:definition3}
For $A, B \in M_{n,sa}(\mathbb{C})$, $\rho \in M_{n,+,1}(\mathbb{C})$ and $f \in {\cal F}_{op}^r$, we define the following quantities:
$$
Corr_{\rho}^f(A,B) \equiv \frac{f(0)}{2} \langle i[\rho,A],i[\rho,B] \rangle_{\rho,f}, \,\, I_{\rho}^f(A) \equiv Corr_{\rho}^f(A,A), 
$$
$$
C_{\rho}^f(A,B) \equiv Tr[m_f(L_{\rho},R_{\rho})(A) B], \,\, C_{\rho}^f(A) \equiv C_{\rho}^f(A,A),
$$
$$
U_{\rho}^f(A) \equiv \sqrt{V_{\rho}(A)^2-(V_{\rho}(A)-I_{\rho}^f(A))^2}.
$$

\end{Def}
The quantity $I_{\rho}^f(A)$ is known as metric adjusted skew information \cite{Han}.
It is notable that  the metric adjusted correlation measure $Corr_{\rho}^c(A,B)$ was 
firstly introduced in \cite{Han} for a regular Morozova-Chentsov function $c$.
Recently the notation $I_{\rho}^c(A,B)$ in \cite{ACH} and  the notation $I_{\rho}^f(A,B)$ in \cite{GI2} were used.
In addition, it is useful for the readers to be noted that 
the correlation $I_{\rho}^f(A,B)$ can be expressed as a difference of covariances \cite{GI2}.
Throughout the present paper, we use the notation $Corr_{\rho}^f(A,B)$ as the metric adjusted correlation measure,
to avoid the confusion of the readers. (In the previous sections, we have already used $Corr_{\rho}(A,B)$, $Corr_{\rho,\alpha}(A,B)$ 
and $Corr_{\rho,\alpha,\gamma}(A,B)$ as correlation measures and done $I_{\rho}(H)$ and $I_{\rho,\alpha}(H)$ as skew informations.)
Then we have the following proposition.

\begin{Prop} {\bf (\cite{GII1,GI})} \label{prop:proposition}
For $A, B \in M_{n,sa}(\mathbb{C})$, $\rho \in M_{n,+,1}(\mathbb{C})$ and $f \in {\cal F}_{op}^r$,  
we have the following relations, where we put $A_0\equiv A-Tr[\rho A]I$ and
$B_0\equiv B-Tr[\rho B]I$.
\begin{itemize}
\item[(1)] $I_{\rho}^f(A) = Tr[\rho A_0^2]-Tr[m_{\tilde{f}}(L_{\rho},R_{\rho})(A_0) A_0] = V_{\rho}(A)-C_{\rho}^{\tilde{f}}(A_0)$.
\item[(2)] $J_{\rho}^f(A) = Tr[\rho A_0^2]+Tr[m_{\tilde{f}}(L_{\rho},R_{\rho})(A_0) A_0] = V_{\rho}(A)+C_{\rho}^{\tilde{f}}(A_0)$.
\item[(3)] $0 \leq I_{\rho}^f(A) \leq U_{\rho}^f(A) \leq V_{\rho}(A)$.
\item[(4)] $U_{\rho}^f(A) = \sqrt{I_{\rho}^f(A) J_{\rho}^f(A)}$.
\item[(5)] $Corr_{\rho}^f(A,B) = \frac{1}{2}Tr[\rho A_0B_0] +  \frac{1}{2}Tr[\rho B_0A_0]  -Tr[m_{\tilde{f}}(L_{\rho},R_{\rho})(A_0) B_0] = \frac{1}{2}Tr[\rho A_0B_0] +  \frac{1}{2}Tr[\rho B_0A_0]  -C_{\rho}^{\tilde{f}}(A_0,B_0)$.
\end{itemize}
\end{Prop}
%{\it Proof}: See \cite{GI} for (1)-(4). (5) follows by the direct calculation. \hfill \qed

The following inequality is the further generalization of Corollary \ref{cor_two_para} by the use of the metric adjusted correlation measure.

\begin{The}\label{th:theorem2}
For $f \in {\cal F}_{op}^r$,  if we have
\begin{equation}
\frac{x+1}{2}+\tilde{f}(x) \geq 2f(x), 
\label{eq:num4-1}
\end{equation}
then we have
\begin{equation}
U_{\rho}^f(A) U_{\rho}^f(B) \geq 4 f(0)|Corr_{\rho}^f(A,B)|^2, 
\label{eq:num4-2}
\end{equation}
for $A, B \in M_{n,sa}(\mathbb{C})$ and $\rho \in M_{n,+,1}(\mathbb{C})$.
\end{The}

In order to prove Theorem \ref{th:theorem2}, we use the following two lemmas. 

\begin{Lem} {\bf (\cite{Yanagi3})}\label{lem:lemma1}
If Eq.(\ref{eq:num4-1}) is satisfied, then we have the following inequality:
$$
\left( \frac{x+y}{2} \right)^2 - m_{\tilde{f}}(x,y)^2 \geq f(0)(x-y)^2.
$$
\end{Lem}

%\begin{flushleft}
{\it Proof}:  By Eq.(\ref{eq:num4-1}), we have
%\end{flushleft} 
$$
\frac{x+y}{2}+m_{\tilde{f}}(x,y) \geq 2 m_f(x,y).
$$
We also have
\begin{eqnarray*}
m_{\tilde{f}}(x,y) & = & y \tilde{f}\left( \frac{x}{y} \right) \\
& = & \frac{y}{2} \left\{ \frac{x}{y}+1-\left( \frac{x}{y}-1 \right)^2 \frac{f(0)}{f(x/y)} \right\} \\
& = & \frac{x+y}{2}-\frac{f(0)(x-y)^2}{2m_f(x,y)}. 
\end{eqnarray*}
Therefore 
\begin{eqnarray*}
\left(\frac{x+y}{2}\right)^2 - m_{\tilde{f}}(x,y)^2 
& = & \left\{ \frac{x+y}{2}-m_{\tilde{f}}(x,y) \right\} \left\{ \frac{x+y}{2}+m_{\tilde{f}}(x,y) \right\} \\
& \geq & \frac{f(0)(x-y)^2}{2 m_f(x,y)} 2m_f(x,y) \\
& = & f(0)(x-y)^2.
\end{eqnarray*}

\hfill \qed

We have the following expressions for the quantities $I_{\rho}^f(A)$, $J_{\rho}^f(A)$, $U_{\rho}^f(A)$ and $Corr_{\rho}^f(A,B) $
by using Proposition \ref{prop:proposition} and a mean $m_{\tilde{f}}$.
\begin{Lem}{\bf (\cite{GI})}  \label{lem:lemma2}
Let $\{ |\phi_1 \rangle, |\phi_2 \rangle, \cdots, |\phi_n \rangle \}$ be a basis of eigenvectors of $\rho$, 
corresponding to the eigenvalues $\{ \lambda_1,\lambda_2,\cdots,\lambda_n \}$. We put 
$a_{jk} = \langle \phi_j|A_0|\phi_k \rangle, b_{jk} = \langle \phi_j|B_0|\phi_k \rangle$, where $A_0\equiv A-Tr[\rho A]I$ and
$B_0\equiv B-Tr[\rho B]I$ for $A, B \in M_{n,sa}(\mathbb{C})$ and $\rho \in M_{n,+,1}(\mathbb{C})$.  Then we have
\begin{eqnarray*}
I_{\rho}^f(A) & = & \frac{1}{2} \sum_{j,k} (\lambda_j+\lambda_k) a_{jk} a_{kj} - \sum_{j,k} m_{\tilde{f}}(\lambda_j,\lambda_k) a_{jk} a_{kj} \\
& = & 2 \sum_{j<k} \left\{ \frac{\lambda_j+\lambda_k}{2}-m_{\tilde{f}}(\lambda_j,\lambda_k) \right\}|a_{jk}|^2,
\end{eqnarray*}
\begin{eqnarray*}
J_{\rho}^f(A) & = & \frac{1}{2} \sum_{j,k} (\lambda_j+\lambda_k) a_{jk} a_{kj} + \sum_{j,k} m_{\tilde{f}}(\lambda_j,\lambda_k) a_{jk} a_{kj} \\
& \geq & 2 \sum_{j<k} \left\{ \frac{\lambda_j+\lambda_k}{2}+m_{\tilde{f}}(\lambda_j,\lambda_k) \right\}|a_{jk}|^2,
\end{eqnarray*}
$$
U_{\rho}^f(A)^2 = \frac{1}{4} \left( \sum_{j,k} (\lambda_j+\lambda_k) |a_{jk}|^2 \right)^2 - \left( \sum_{j,k} m_{\tilde{f}}(\lambda_j,\lambda_k) |a_{jk}|^2 \right)^2
$$
and
\begin{eqnarray}
Corr_{\rho}^f(A,B) & = & \frac{1}{2} \sum_{j,k} \lambda_j a_{jk}b_{kj}  +\frac{1}{2} \sum_{j,k} \lambda_k a_{jk}b_{kj}   -\sum_{j,k} m_{\tilde{f}}(\lambda_j,\lambda_k) a_{jk}b_{kj} \nonumber \\
& = & \sum_{j<k} \left(  \frac{\lambda_j+\lambda_k}{2}  -m_{\tilde{f}}(\lambda_j,\lambda_k)\right)a_{jk}b_{kj} + \sum_{j<k} \left(\frac{\lambda_k+\lambda_j}{2}-m_{\tilde{f}}(\lambda_k,\lambda_j)\right)a_{kj}b_{jk}.\nonumber \\
 \label{eq_corr_f_expression}
\end{eqnarray}
\end{Lem}

We are now in a  position to prove Theorem \ref{th:theorem2}.

%\begin{flushleft}
{\it Proof of Theorem \ref{th:theorem2}}:
From Eq.(\ref{eq_corr_f_expression}), we have 
%\end{flushleft}
\begin{eqnarray*}
 |Corr_{\rho}^f(A,B)| & \leq & \sum_{j<k} \left|\left(\frac{\lambda_j+\lambda_k}{2}  -m_{\tilde{f}}(\lambda_j,\lambda_k)\right)a_{jk}b_{kj}\right|
 + \sum_{j<k} \left|\left(\frac{\lambda_j+\lambda_k}{2}  -m_{\tilde{f}}(\lambda_k,\lambda_j)\right)a_{kj}b_{jk}\right| \\
& \leq & \sum_{j<k} \left|\frac{\lambda_j+\lambda_k}{2}  -m_{\tilde{f}}(\lambda_j,\lambda_k)\right| |a_{jk}| |b_{kj}|
 + \sum_{j<k} \left|\frac{\lambda_j+\lambda_k}{2}  -m_{\tilde{f}}(\lambda_k,\lambda_j)\right| |a_{kj}| |b_{jk}| \\
%& = & \sum_{j<k} \left(\left| \frac{\lambda_j+\lambda_k}{2}  -m_{\tilde{f}}(\lambda_j,\lambda_k)\right|
%+\left|\frac{\lambda_j+\lambda_k}{2}  -m_{\tilde{f}}(\lambda_k,\lambda_j)\right|\right) |a_{jk}| |b_{kj}| \\
& = &  2 \sum_{j<k} \left| \frac{\lambda_j+\lambda_k}{2}  -m_{\tilde{f}}(\lambda_j,\lambda_k) \right|  |a_{jk}| |b_{kj}| \\
& \leq & \sum_{j<k} |\lambda_j-\lambda_k| |a_{jk}| |b_{kj}|. 
\end{eqnarray*}
Then we have 
\begin{eqnarray*}
 f(0) | Corr_{\rho}^f(A,B)|^2
& \leq & \left( \sum_{j<k} f(0)^{1/2} |\lambda_j-\lambda_k| |a_{jk}| |b_{kj}| \right)^2 \\
& \leq & \left( \sum_{j<k} \left\{ \left( \frac{\lambda_j+\lambda_k}{2} \right)^2 - m_{\tilde{f}}(\lambda_j,\lambda_k)^2 \right\}^{1/2} |a_{jk}| |b_{kj}| \right)^2 \\
& \leq & \left( \sum_{j<k} \left\{ \frac{\lambda_j+\lambda_k}{2} - m_{\tilde{f}}(\lambda_j,\lambda_k) \right\} |a_{jk}|^2 \right) \\
&   &  \times \left( \sum_{j<k} \left\{ \frac{\lambda_j+\lambda_k}{2} + m_{\tilde{f}}(\lambda_j,\lambda_k) \right\} |b_{kj}|^2 \right) \\
& \leq & \frac{1}{4}I_{\rho}^f(A) J_{\rho}^f(B).
\end{eqnarray*}
By the similar way, we also have
$$
I_{\rho}^f(B) J_{\rho}^f(A) \geq 4 f(0) |Corr_{\rho}^f(A,B)|^2.
$$
Hence we have the desired inequality (\ref{eq:num4-2}). 

\hfill \qed

\begin{Rem}
Under the same assumptions with Theorem \ref{th:theorem2}, we have the following Heisenberg-type uncertainty relation \cite{Yanagi3}:
\begin{equation} \label{rem4_10}
U_{\rho}^f(A) U_{\rho}^f(B) \geq  f(0) |Tr\left[\rho[A,B]\right]|^2
\end{equation}
by the similar way to the proof of Theorem \ref{th:theorem2}, since we have
$$
 |Tr\left[\rho[A,B]\right]| \leq 2 \sum_{j<k}\vert \lambda_j -\lambda_k\vert \vert a_{jk}\vert \vert b_{kj} \vert.
$$ 
As stated in Remark \ref{rem_corr_tr}, there is no ordering between the right hand side of the inequality (\ref{eq:num4-2}) and that of the inequality (\ref{rem4_10}), in general.
\end{Rem}
%\vspace{0.5cm}
%\noindent

If we use the function 
$$
f_{WYD}(x) = \alpha (1-\alpha) \frac{(x-1)^2}{(x^{\alpha}-1)(x^{1-\alpha}-1)},\quad \alpha \in (0,1),
$$
then we obtain the following uncertainty relation.

\begin{Cor}
For $A, B \in M_{n,sa}(\mathbb{C})$ and $\rho \in M_{n,+,1}(\mathbb{C})$, we have 
$$
U_{\rho}^{f_{WYD}}(A) U_{\rho}^{f_{WYD}}(B) \geq 4\alpha(1-\alpha)|Corr_{\rho}^{f_{WYD}}(A,B)|^2.
$$
\label{cor:corollary1}
\end{Cor}

%\begin{flushleft}
{\it Proof}: From the definition 
%\end{flushleft}
$$
f_{WYD}(x) = \alpha(1-\alpha) \frac{(x-1)^2}{(x^{\alpha}-1)(x^{1-\alpha}-1)},
$$
it is clear that 
$$
\tilde{f}_{WYD}(x) = \frac{1}{2} \{ x+1-(x^{\alpha}-1)(x^{1-\alpha}-1) \}. 
$$
By Lemma \ref{lem-1-2-02}, we have for $0 \leq \alpha \leq 1$ and $x > 0$, 
$$
(1-2\alpha)^2(x-1)^2-(x^{\alpha}-x^{1-\alpha})^2 \geq 0.
$$
This inequality can be rewritten by
$$
(x^{2\alpha}-1)(x^{2(1-\alpha)}-1) \geq 4 \alpha (1-\alpha)(x-1)^2.
$$
Thus  we have
\begin{eqnarray*}
 \frac{x+1}{2}+\tilde{f}_{WYD}(x) 
& = & x+1-\frac{1}{2}(x^{\alpha}-1)(x^{1-\alpha}-1) \\
& = & \frac{1}{2}(x^{\alpha}+1)(x^{1-\alpha}+1) \\
& \geq & 2 \alpha (1-\alpha) \frac{(x-1)^2}{(x^{\alpha}-1)(x^{1-\alpha}-1)} \\
& = & 2 f_{WYD}(x). 
\end{eqnarray*}
Thus we obtain the aimed result from Theorem \ref{th:theorem2}. 

\hfill \qed

Note that Corollary \ref{cor_two_para} coincides with  Corollary \ref{cor:corollary1}, since we have $U_{\rho,\alpha}(A)=U_{\rho}^{f_{WYD}}(A) $
which is obtained by the fact the function $f_{WYD}(x)$ corresponds to the Wigner-Yanase-Dyson skew information.
We also note that we have $Corr_{\rho}^{f_{WYD}}(A,B) = Corr^{(sym)}_{\rho,\alpha,\frac{1}{2}}(A,B)$  and \\ $Corr_{\rho}^{f_{WYD}}(A,B)\neq Corr_{\rho,\alpha,\frac{1}{2}}(A,B)$ in general.

%%%%%%%%%%%%%%%%%%%%%%%%%%%%%%%%%%%%%%%%%%%%%%%%%%%%%%%%%%%%%%%%%%%%%%%%%%%%%%%%%%%%%%%%%%%%%%%%%%%%%%%%%%%%%%%%%%%%%%%%%%%%%%%%%%%%%%%%%%%%
%%%%%%%%%%%%%%%%%%%%%%%%%%%%%%%%%%%%%%%%%%%%%%%%%%%%%%%%%%%%%%%%%%%%%%%%%%%%%%%%%%%%%%%%%%%%%%%%%%%%%%%%%%%%%%%%%%%%%%%%%%%%%%%%%%%%%%%%%%
%%%%%%%%%%%%%%%%%%%%%%%%%%%%%%%%%%%%%%%%%%%%%%%%%%%%%%%%%%%%%%%%%%%%%%%%%%%%%%%%%%%%%%%%%%%%%%%%%%%%%%%%%%%%%%%%%%%%%%%%%%%%%%%%%%%%%%%%%%
%%%%%%%%%%%%%%%%%%%%%%%%%%%%%%%%%%%%%%%%%%%%%%%%%%%%%%%%%%%%%%%%%%%%%%%%%%%%%%%%%%%%%%%%%%%%%%%%%%%%%%%%%%%%%%%%%%%%%%%%%%%%%%%%%%%%%%%%
%%%%%%%%%%%%%%%%%%%%%%%%%%%%%%%%%%%%%%%%%%%%%%%%%%%%%%%%%%%%%%%%%%%%%%%%%%%%%%%%%%%%%%%%%%%%%%%%%%%%%%%%%%%%%%%%%%%%%%%%%%%%%%%%%%%%%%%%
%%%%%%%%%%%%%%%%%%%%%%%%%%%%%%%%%%%%%%%%%%%%%%%%%%%%%%%%%%%%%%%%%%%%%%%%%%%%%%%%%%%%%%%%%%%%%%%%%%%%%%%%%%%%%%%%%%%%%%%%%%%%%%%%%%%%%%%%%%

\section*{Acknowledgements}
The authors thank to anonymous referees for giving us valuable comments and suggestions to improve our manuscript. 
The authors also thank to Dr.F.C.Mitroi for giving us valuable comments to improve our manuscript. 
The author (S.F.) was partially supported by the Japanese Ministry of Education, Science, Sports and Culture,
Grant-in-Aid for Encouragement of Young Scientists (B), 20740067.  Also the author (K.Y.) was partially supported by the Japanese Ministry of Education, Science, 
Sports and Culture, Grant-in-Aid for Scientific Research (C), 23540208.

\end{document}